\documentclass[12pt,a4paper,superscriptaddress,preprint]{revtex4-1}
\usepackage[dvips]{graphicx}
\usepackage{amssymb,amsmath}

\usepackage{dcolumn}% Align table columns on decimal point
\usepackage{bm}% bold math
\usepackage{booktabs}
\usepackage{type1cm}
\newcommand{\average}[1]{\ensuremath{\langle#1\rangle} }

\begin{document}

\title{Universal bursty behavior in the air transportation system}% Force line breaks with \\

\author{Hidetaka Ito}
\email{ito@jamology.rcast.u-tokyo.ac.jp}
\affiliation{%
Department of Aeronautics and Astronautics, School of Engineering, The University of Tokyo,\\
7-3-1 Hongo, Bunkyo-ku, Tokyo 113-8656, Japan
}%
\thanks{}%
\author{Katsuhiro Nishinari}%
\affiliation{%
Research Center for Advanced Science and Technology, The University of Tokyo,\\
4-6-1 Komaba, Meguro-ku, Tokyo 153-8904, Japan
}%

\date{\today}% It is always \today, today,

\begin{abstract}
Social activities display bursty behavior characterized by heavy-tailed inter-event time distributions. We examine the bursty behavior of airplanes' arrivals in hub airports. The analysis indicates that the air transportation system universally follows a power-law inter-arrival time distribution with an exponent $\alpha=2.5$ and an exponential cutoff. Moreover, we investigate the mechanism of this bursty behavior by introducing a simple model to describe it. In addition, we compare the extent of the hub-and-spoke structure and the burstiness of various airline networks in the system. Remarkably, the results suggest that the hub-and-spoke network of the system and the carriers' strategy to facilitate transit are the origins of this universality.

\end{abstract}

\maketitle

%\tableofcontents

\section{INTRODUCTION}
\label{sec:1}
Considerable attention has been paid to the dynamics of social activities and networks \cite{RevModPhys.74.47, RevModPhys.81.591, arenas2008synchronization, barthelemy2011spatial, holme2012temporal}. The availability of a large amount of data has recently enabled researchers to analyze the detailed structures of social activity patterns. Burstiness has recently been considered as a fundamental pattern of social phenomena: the probability density functions (PDFs) of inter-event times (IETs) of many social activities are characterized by heavy tails. This is evidence of the non-Poissonian nature of social activities, which indicates that each social activity is strongly correlated with other activities. The heavy-tailed structure of the IET distribution is well-approximated by a power-law tail with an exponential cutoff $p(\tau) = e^{-\tau/\tau_0}\tau^{-\alpha}$, where $\alpha$ is an exponent of the power law \cite{eckmann2004entropy, karsai2012universal}. This structure is universally observed in various phenomena including human activities, such as sending emails and library loans \cite{eckmann2004entropy, vazquez2006modeling, goh2008burstiness, candia2008uncovering, takaguchi2011predictability, cattuto2010dynamics}, and natural phenomena, such as a neuron's interspike \cite{kemuriyama2010power} and an earthquake's shock intervals \cite{saichev2006universal}. Furthermore, the bursty behaviors of systems have a strong influence on the collective phenomena in their networks \cite{PhysRevLett.114.108701, iribarren2009impact, gavalda2014impact, horvath2014spreading}. The effect of bursts spreading processes on networks, has recently been studied empirically, numerically, and analytically \cite{10.1371/journal.pone.0068629, vazquez2007impact2, karsai2011small, jo2014analytically}. These studies indicate that burstiness is a significant factor in understanding social phenomena on networks.

Moreover, proposing reasonable explanations and models for these activities has been a significant issue \cite{hidalgo2006conditions}. One such model describing bursty behavior is a queueing model in which an individual prioritizes some important tasks based on the assumption that humans have a wide range of tasks and attempt to deal with the urgent ones immediately \cite{barabasi2005origin, vazquez2005exact, vazquez2006modeling, vajna2013modelling}. Another possible explanation is a cascading Poisson process with a circadian rhythm. Once one engages in an action such as sending an e-mail, one continuously repeats the action for a while although the initiation of the actions is independent of other actions; this is called a cascading non-homogeneous Poisson process. Malmgren {\it et al.} proposed a model in which agents start a cascade of actions at a rate determined by their circadian rhythms \cite{malmgren2008poissonian}. Jo {\it et al.} claimed, based on empirical data analysis, that human mobile phone communication has bursts without a circadian rhythm \cite{jo2012circadian}. Some argue that bursts stem from the memory effect \cite{masuda2013self, vestergaard2014memory, vazquez2007impact1}. However, these possible explanations of bursty behaviors in social phenomena are difficult to apply to components of social systems such as the air transportation system.

The air transportation system has attracted considerable attention because of its importance to mobility. This system consists of hubs and spokes and has small-world and scale-free characteristics \cite{bryan1999hub, guimera2005worldwide, zanin2013modelling}. The assortativity, multiplexity, and epidemic spreading in the network all have been the themes of recent studies \cite{li2004statistical, cardillo2013emergence, colizza2006role}. In addition, constructing resilient air transportation systems is of utmost importance to our society in terms of reliability. The influence of the air transportation network structure on robustness against perturbations has been analyzed \cite{wuellner2010resilience, fleurquin2013systemic}. In addition, the bursty arrival of airplanes is a cause of traffic congestion in the air transportation system \cite{peterson1995models, gwiggner2014data}. This is a significant source of destabilization in the system. It is necessary to study the extent of burstiness to assess its effect on the system. Nevertheless, the burstiness of the air transportation system is not well understood. In particular, understanding the bursty behavior in hub airports is of significance because airplanes' arrivals are concentrated in hub airports due to their small-world characteristic.

Therefore, in this paper, we first analyze the inter-arrival time probability distributions of airplanes in U.S. hub airports. Arrivals of airplanes in each hub airport correspond to events and the inter-arrival time is called the inter-event time (IET) in this analysis. We found that the IET distributions of airplanes in U.S. hub airports follow power-law tails with an exponent $\alpha = 2.5$ and an exponential cutoff. The extent of burstiness is assessed by the cutoff value of the power-law. Next, the origin of the universal bursty behavior is studied. We study the origin of burstiness using a simple model and identify that it originates from airlines' strategy to facilitate transit at hub airports. Moreover, we analyze the relation between each airline's network and the extent of burstiness of the airplanes' arrival behavior in its hub airports. The result indicates that the hub-and-spoke structure of the network is important in the bursty behavior of the system.

The remainder of this paper is organized as follows. In Sec.~\ref{sec:2}, the empirical data of airplanes' arrivals are studied. We reveal that the IET distributions in hub airports are power-law distributions with an exponential cutoff. In Sec.~\ref{sec:3}, we investigate the origin of this universally observed characteristic by proposing a model describing airlines' strategy to facilitate transit. In Sec.~\ref{sec:4}, we discuss the relationship between airline networks and the extent of burstiness in their hub airports. Section~\ref{sec:5} provides to the conclusion.

\section{UNIVERSAL BURSTINESS IN EMPIRICAL AIR TRAFFIC DATA}
\label{sec:2}

\begin{figure}[t]
\begin{center}
\includegraphics[width=7cm]{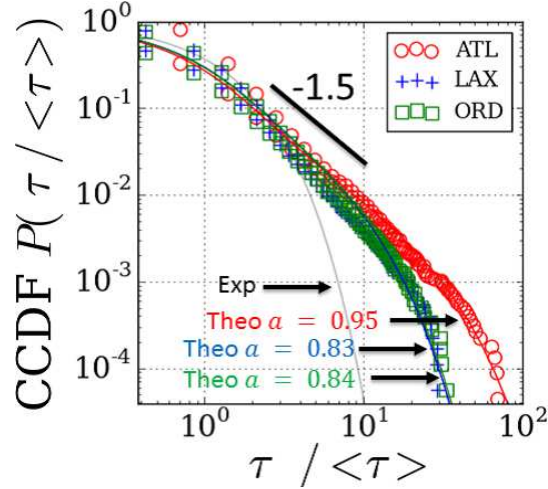}
\caption{(Color online) The CCDFs of the IET of airplanes in three hub airports. The maximum and minimum probability for each IET, $\tau$, are plotted, since many IETs have the same value. The gray line is the exponential distribution. The lines are the theoretical results for the CCDF of the IET of the Sine models for the parameter shown in the figure. All three distributions follow power laws with an exponent $\alpha=2.5$ and an exponential cutoff. The theoretical lines of the Sine models agree well with the empirical data.}
\label{fig:1}
\end{center}
\end{figure}

In this section, we discuss burstiness of airplanes' arrivals using empirical data. We analyze the IET distributions of airplanes in the 10 largest hub airports in the U.S. based on passenger boardings in 2014 (see Appendix A for details on the dataset and data processing). The number of arrivals from all airplanes are counted for each airport. In Fig.~\ref{fig:1}, the CCDFs of the IETs in three major hub airports are shown. In this paper, the horizontal axis is the IET divided by the average of the IET, $\tau / \average{\tau}$. According to the figure, it is universally observed that distributions follow power-laws with an exponent $\alpha=2.5$ and an exponential cutoff as 
\begin{equation}
P(\tau) \sim e^{-\tau/\tau_0} \tau^{-2.5},
\end{equation} 
where $\tau_0$ denotes the cutoff value.

We assess the cutoff value $\tau_0$ as follows: consider a theoretically tractable model following an inhomogeneous Poisson process. Such a process is a system whose events occur at a time-dependent rate $f(t)$, with each event occurring independently \cite{galliher1958nonstationary}. Let us define a model whose event rate is given by 
\begin{equation}
f(t)=N a \sin (2n \pi t) + 1~((0 \le t \le 1)
\end{equation}
as the {\it Sine model}, where $a \in \mathbb{R}$ and $n \in \mathbb{N}$ are parameters and $N$ represents the average number of total events in a trial. The system starts and ends at the times $t=0$ and $t=1$, respectively. Hereafter, we discuss the Sine model taking the limit as $N \to \infty$. In this case, the CCDF of the IET of the Sine model is approximately given by a power-law distribution with an exponent $\alpha = 2.5$ and an exponential cutoff when $\tau$ is large, regardless of parameters other than $a=0$ (constant event rate). $\tau_0$ depends only on the parameter $a$ and is given by $\tau_0 = 1/(1-a)$ (see Appendix F for details). Using this result, the logarithm of the theoretically calculated CCDF of the IET of the Sine model is fitted to the logarithm of the empirical data. The IET distribution of the Sine model is uniquely determined when $a$ is given, since the result is independent of $n$. Then, we obtain fitted parameter $a$ and calculate $\tau_0$. In addition, we introduce $a$ as another metrics for the extent of the burstiness and call it the {\it burst strength parameter}. The parameter $a$ is utilized to assess the extent of burstiness of the IET distributions in terms of the amplitude of the event rate, with $1-a$ representing the minimum event rate. The larger the extent of the burstiness, the larger both the cutoff value and burst strength parameter are.

The theoretically obtained CCDFs of the Sine models for the parameters fitted with the empirical data are also shown in Fig.~\ref{fig:1}.  Theses lines agree well with the empirical data including the cutoff area, which indicates that the assumption of an inhomogeneous Poisson process and an event rate $f(t)=Na \sin (2n \pi t) + 1~(0 \le t \le 1,~N \to \infty)$ is valid for fitting the empirical data to the model. As fitting to the Sine models is an appropriate method for assessing the extent of burstiness in the air transportation system, we utilize this method throughout the paper.

\begingroup
\renewcommand{\arraystretch}{1.2}
\begin{table}[t]
  \begin{tabular}{lcccccc} \toprule \toprule
    Airport & State & $N_{total}$ & Main Carrier & Share & $a$ & $\tau_0$ \\ \hline
    ATL & GA & 28226 & DL & 69.25\% & 0.95 & 19.11 \\
    LAX & CA & 17990 & UA & 18.91\% & 0.83 & 5.95  \\ 
    ORD &  IL & 17840 & UA & 26.81\% & 0.84 & 6.21 \\ 
    DFW & TX & 22701 & AA & 69.23\% & 0.90 & 10.00 \\ 
    JFK & NY & 7059 & B6 & 37.74\% & 0.67 & 2.99 \\ 
    DEN & CO & 17160 & WN & 26.41\% & 0.85 & 6.59 \\ 
    SFO & CA & 13054 & UA & 39.13\% & 0.86 & 7.40 \\ 
    CLT & NC & 9306 & US & 59.25\% & 0.88 & 8.39 \\
    LAS & NV & 10807 & WN & 43.85\% & 0.86 & 7.11 \\
    PHX & AZ & 13051 & US & 26.41\% & 0.86 & 6.43 \\ \bottomrule \bottomrule
  \end{tabular}
\caption{The burst strength parameter, $a$, and the cutoff value, $\tau_0$, of the power-law distribution of 10 major airports. $N_{total}$ denotes the number of total arrivals in the dataset. Although the locations and number of arrivals, and the main carrier of these airports are different, most airports have large cutoff values, which indicates that the bursty behaviors of airplanes' arrivals in hub airports are universally observed.}
\label{Table:1}
\end{table}
\endgroup

The burst strength parameter, $a$, and the cutoff value, $\tau_0$, of the power-law distribution of 10 hub airports are shown in Table~\ref{Table:1}. Each airport's IATA code, the state in which it is located, its number of arrivals, its main carrier, and its main carrier's share are also shown. The table shows that the cutoff value of airports located in a wide range of areas in the U.S. is large, which indicates that the IET distribution of arrivals in these airports have bursty behaviors. The extent of burstiness in these airports is universally large, although they have differences in terms of the main-airline-operating airplanes, locations, and passengers' destinations (each airport's cutoff value is slightly different; see Appendix B for the analysis of this difference in the cutoff value among hub airports.) This clearly shows that the mechanism generating this universal bursty behavior exists in the air transportation system.

\section{ORIGIN OF BURSTINESS IN THE AIR TRANSPORTATION SYSTEM}
\label{sec:3}

\subsection{Description of transit facilitation}

\begin{figure}[t]
\begin{center}
\includegraphics[width=7cm]{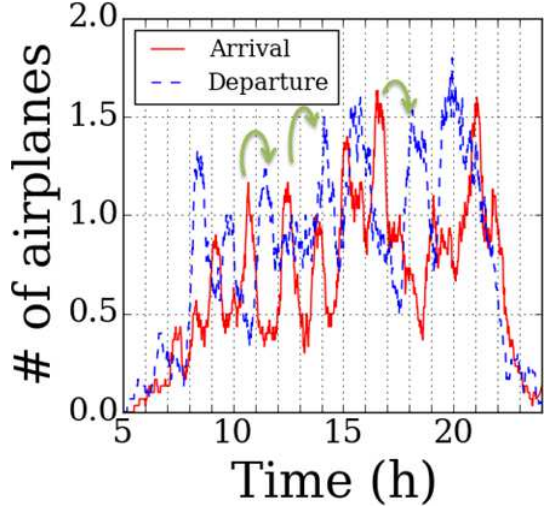}
\caption{(Color online) The 30-min moving averages of the numbers of arrivals and departures in a day. Thirty minutes is the time interval during which most passengers make a plane connection. Heterogeneity of the arrival and departure behaviors contributes to transit facilitation. Arrows represent examples of transit plans making the most of the transit facilitation strategies of airlines.}
\label{fig:3}
\end{center}
\end{figure}

In this section, we discuss the origin of the universally observed bursty behavior. The mechanism behind the bursty behavior in the air transportation system is different from the explanations of bursty behavior in other systems. First, we consider a factor that affects the scheduled arrival time of each airplane to model airplanes' arrival dynamics. Facilitation of passengers' transit at airports is the main factor affecting the flight schedule. Transit plays an important role in the air transportation system. Pan {\it et al.} showed that temporal distances for the air transportation network are shorter than the {\it time-shuffled model}, in which the time stamps of all arrivals are shuffled \cite{pan2011path}. The schedule of the system is optimized to efficiently transport passengers. The facilitation of a passenger's transit is an airline's traffic optimization strategy, which shortens the temporal distance. 

Let us discuss the necessity of transit facilitation strategy of airlines. The air transportation network is composed of hubs and spokes. This network enables passengers to travel to a variety of destinations via hub airports because of their small-world characteristic. Passengers arriving at these airports transfer from airplanes to various other airplanes. These airports have a strong demand for facilitating the transit of passengers. Thus, airlines try to present more destinations choices to their passengers. Airlines have to prevent passengers from accidents such as transit failures because of the very short time for transit, while fewer passengers choose a connecting flight whose departure time is very long after the arrival of the passengers' previous flight. 

To achieve these conditions, the following facilitation strategy is adopted: all possible airplanes that transit passengers might board are arranged to arrive at the airport at almost the same time. In addition, connecting flights also depart the airport at almost the same time. Let us call these times {\it arrival/departure-concentrated times}. The time interval for transit is long enough that passengers can have fewer accidents due to a shorter transit time and short enough that passengers will choose the flight as a connecting flight. These combinations cause passengers' transit times to remain almost constant regardless of their sources and destinations, thereby facilitating transit.

\begin{figure*}
\begin{center}
\includegraphics[width=16cm]{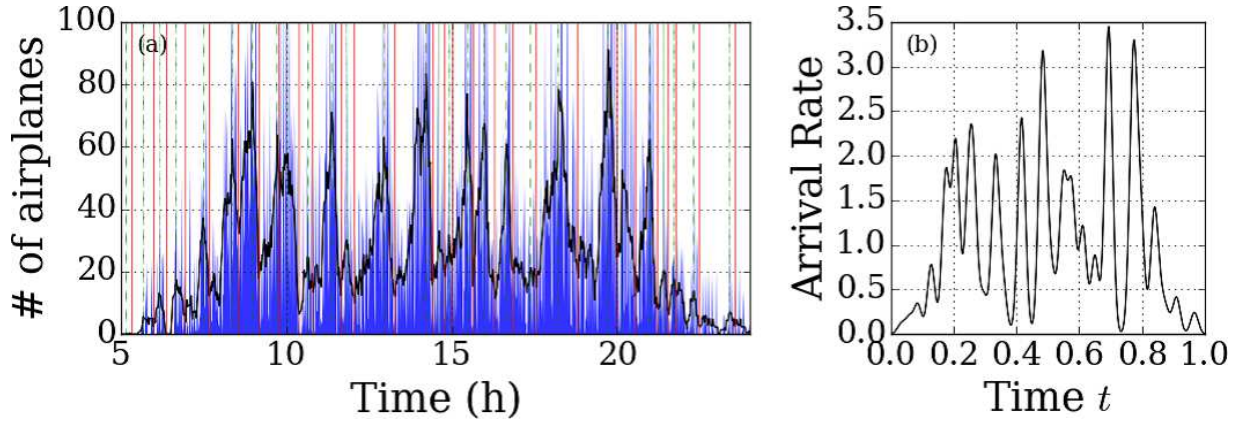}
\caption{(Color online) (a) Histogram of scheduled arrival airplanes and their 5-min moving average are shown in blue (gray in gray-scale) and black, respectively. Solid and dashed vertical lines are local minimums and maximums, respectively. The area is divided into regions by the local minimums. As an approximation, airplanes' arrivals are concentrated on the local maximums in each region for constructing the Normal distribution model. (b) Arrival rate of the Normal distribution model constructed using the empirical data of ATL.}
\label{fig:4}
\end{center}
\end{figure*}

We discuss the influence of facilitation of passengers' transit on flight schedule. Several flights are scheduled to arrive at the airport at the arrival-concentrated time, while not many airplanes are expected at other times. This factor causes airports to fluctuate the number of arrivals during a certain time span. The number of arrivals as a function of time, $t$, has many local peaks. In Fig.~\ref{fig:3}, the 30-min moving average of the number of arrivals and departures on January 31, 2014 is shown. Since most passengers take 45--75 min to make a plane connection \cite{barnhart2014modeling}, the 30-min moving average of the number of departure flights at time $t$ indicates the number of flights with which passengers arriving at the airport 60 min before time are able to make a connection. The figure shows that the number of arrivals and departures alternately approach their local peaks with the passage of time. Arrows are instances of transit plans that allows passengers to benefit from various options for connecting flights. For instance, passengers who arrive at the airport at around 16:30 can comfortably leave the airport at around 18:00 because of transit facilitation.

\subsection{Normal distribution model}

Next, to understand the effect of facilitation of passengers' transit on burstiness in the air transportation system, we propose a simple model called the {\it Normal distribution model}, which shows bursty behavior. The Normal distribution model is constructed as follows. As with the Sine model, we assume that the model follows an inhomogeneous Poisson process, whose arrival rate is given by a function of time, $f(t)$. First, we mention the scheduled arrival rate $f_{schedule}(t)$ and then we remark on the actual arrival rate$f(t)$, with considering the schedule and the delay. 

The scheduled arrival rate has local peaks at the arrival-concentrated time, which describes the transit facilitation strategy mentioned above. Let $\mu_i$ denote the $i$th arrival concentrated time. We assume that these peaks of the scheduled arrival rate can be modeled by the delta function $\delta (t-\mu_i)$. The scheduled arrival rate is $0$ except at the arrival-concentrated time. This is the extreme case of the concentration of airplanes' arrivals. This assumption is valid when the peakedness of these peaks is high enough. The scheduled arrival rate is given by $f_{schedule}(t) = \sum_{i} c_i \delta (t-\mu_i)~(i=1, \ldots, n)$, where $c_i$ are constants. 

Let us discuss the actual arrival rate $f(t)$ of the Normal distribution model. We consider the effect of randomness upon modeling. Although pilots aim to reach the destination at the scheduled time, the actual arrival time is delayed by randomness based on the weather, other airplanes, and so forth. Negative delay times indicate early arrivals. The arrival delay time distribution can be modeled by a normal distribution with a mean of $-2.73$ min and a standard deviation of 13.75 min according to analysis of the empirical data \cite{mueller2002analysis}. Thus, we assume that the delay time distribution is given by the normal distribution $\mathcal{N}(\mu_{delay}, \sigma^2)$ where $\mu_{delay} = -2.73~{\rm min.}$ and $\sigma = 13.75~{\rm min.}$ The actual arrival time is spread out following normal distribution. Considering the scheduled arrival rate and delay distribution, the actual arrival rate is given by the mixture of the normal distributions
\begin{equation}
\label{sigma}
f(t) = \sum_{i} \frac{c_i}{\sigma \sqrt{2 \pi}} e^{-(x-\bar{\mu_i})^2/2 \sigma^2},
\end{equation}
where $\bar{\mu_i} = \mu_i + \mu_{delay}$.

Let us mention the fitting process of the Normal distribution model to the empirical data. We fit the empirical data for the amount of scheduled arrivals to delta functions $f_{schedule}(t)$, and then calculate the actual arrival rate $f(t)$. First, we divide the time space of the empirical data into subregions and set a peak in each subregion (see Appendix C for details). In Fig.~\ref{fig:4}(a), the region segmentation and peak setting results for the case of ATL in January 2014 is shown. The histogram of scheduled arrivals and the 5-min average of arrivals are shown in blue (gray in gray-scale) and black, respectively. The time space is divided into subregions by vertical solid lines. The peak in each subregion is represented by a vertical dashed line. Then, the number of scheduled arrivals in each subregion, $N_i$, is counted. In fitting, we assume that all scheduled arrivals concentrated on the peak of each subregion. Thus, the height of each peak, $c_i$, is proportional to $N_i$. The actual arrival rate is obtained by substituting $c_i$ into Eq.~\ref{sigma} for the fitted value by considering the normalization condition $\int_0^1 f(t) dt = 1$. The fitting result of the arrival rate of the Normal distribution model using the empirical data in the ATL case is shown in Fig.~\ref{fig:4}(b). The time is rescaled to $0 \le t \le 1$.

\begin{figure}[t]
\begin{center}
\includegraphics[width=7cm]{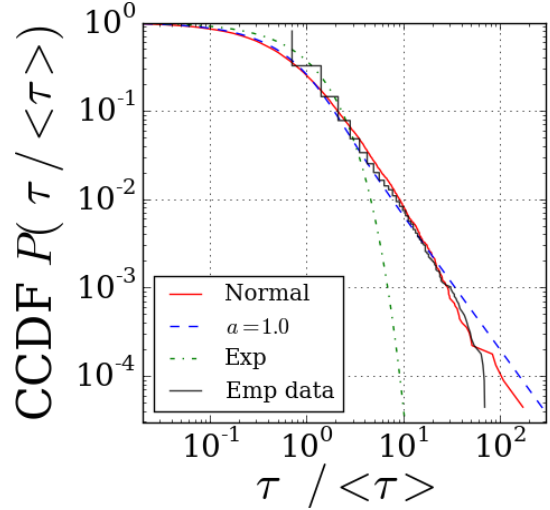}
\caption{(Color online) Comparison among the CCDFs of the IET of the Normal distribution model constructed with empirical data of the arrival behavior in ATL, Sine model $(a = 1.0)$, (time-independent) Poisson process, and empirical data in the ATL case. The IET distribution of the Normal distribution model is in agreement with the empirical data and follows a power-law distribution with an exponent $\alpha= 2.5$ and a cutoff.}
\label{fig:5}
\end{center}
\end{figure}

\subsection{Results of model analysis}

Let us discuss the simulation result for the Normal distribution model. In Fig~\ref{fig:5}, the CCDF of the IET of the Normal distribution model constructed using the empirical data in the case of ATL is shown. The IET distribution of this model is compared with those of the empirical data of arrivals in ATL and the Sine models for the parameters $a=0.0, 1,0$. The IET distribution of the Normal distribution model follows a power law with an exponent $\alpha = 2.5$ and an exponential cutoff. In addition, the IET distribution agrees well with the empirical data. The cutoff values of the power law and burst strength parameters of the model are $\tau_0 = 28.07$ and $a = 0.97$, respectively, which are similar to those of the empirical data, $\tau_0 = 19.11$ and $a=0.95$.

We theoretically discuss the CCDF of the IET of this model. There is a local minimum, $t = \tilde{t_i}$, between two consecutive peaks of normal distributions in Fig.~\ref{fig:4}(b). The event rate, $f(t)$, can be expanded as $f(t - \tilde{t_i}) = c_{1i} + \mathcal{O}((t-\tilde{t_i})^2)$ asymptotically at this point given that $\mu_{i+1} - \mu_i$ is sufficiently small compared with $\sigma_i$. If the event rate is a quadratic function, the CCDF of the IET follows a power-law tail with an exponent $\alpha = 2.5$ \cite{vazquez2007impact1}. Thus, the Normal distribution model can generate this IET distribution. This result indicates that the assumption of the Normal distribution model that transit facilitation is a factor affecting the airplanes' arrival behavior is indeed the fundamental mechanism of the bursty behavior in the air transport system.

The universality of the bursty behavior originates from the robustness of this mechanism against variations in the scheduled arrival time distribution; the reason for this robustness is discussed. The peakedness of the peak of the arrival rate distribution is notasso high as a delta function because very high traffic concentrations should be avoided because of limited traffic capacity. In addition, the arrival delay time distribution is asymmetric \cite{lan2006planning}. These two factors affect the scheduled and actual arrival time distributions. However, the exponent of the power law in the IET distribution of the model remains 2.5 as long as the delay time distribution is given by a smooth function since the event rate is expanded as a quadratic function at the local minimum points in this case. In addition, time intervals between two consecutive peaks in the event rate does not greatly affect burstiness for the same reason.

\begingroup
\renewcommand{\arraystretch}{1.2}
\begin{table*}[t]
\begin{center}
\begin{tabular}{lccccccccccccc} \toprule \toprule
Carrier & $N$ & $E$ & $S$ & $\average{q}$ & $\average{s}$ & $\average{l}$ & $\average{C}$ & $r$ & $G(q)$ & $G(s)$ & Airport & $a$ & $\tau_0$\\ \hline
AA & 84 & 354  & 43711& 8.43 & 1041 & 2.01  & 0.53  & -0.64  & 0.62  & 0.70  & DFW & 0.92 & 12.63 \\
DL & 134 & 650  & 55928 & 9.70 & 835 & 2.11  & 0.44  & -0.58  & 0.66  & 0.76  & ATL & 0.87 & 7.91 \\
US & 81 & 315 & 33651 & 7.78 & 831 & 2.15  & 0.50  & -0.82  & 0.60  & 0.73  & CLT & 0.86 & 7.05 \\
AS & 54 & 207 & 12169 & 7.67 & 451 & 2.27  & 0.30  & -0.48  & 0.53  & 0.64  & SEA & 0.76 & 4.15 \\
UA & 81 & 497 & 37291 & 12.27 & 921 & 2.15  & 0.58  & -0.72  & 0.62  & 0.76  & ORD & 0.70 & 3.29 \\
WN & 89 & 1078 & 86698 & 24.22 & 1948 & 2.00  & 0.66  & -0.48  & 0.51  & 0.59  & DEN & 0.57 & 2.31 \\
B6 & 55 & 276 & 17966 & 10.04 & 653 & 2.11  & 0.54  & -0.56  & 0.54  & 0.63  & JFK & 0.40 & 1.67 \\ \bottomrule \bottomrule
\end{tabular}
\caption{The properties of airlines' networks, the burst strength parameter, $a$, and the cutoff value, $\tau_0$, of the power-law distribution in these airlines' main hub airports. $N$, $E$, and $S$ denote the number of nodes, edges without multiple edges, and edges with multiple edges, respectively. $\average{q}$ and $\average{s}$ denote the average node degree and strength, respectively. $\average{l}$, $\average{C}$, and $r$ denote the average shortest path length, average clustering coefficient, and degree assortativity, respectively. $G(q)$ and $G(s)$ denote the degree and strength Gini coefficients, respectively. FSCs and LCCs are characterized by small and large cutoff values, respectively.}
\label{Table:2}
\end{center}
\end{table*}
\endgroup

\section{RELATIONSHIP BETWEEN AIRLINE NETWORKS AND THEIR BURSTY BEHAVIORS}
\label{sec:4}

\begin{figure}[t]
\begin{center}
\includegraphics[width=7cm]{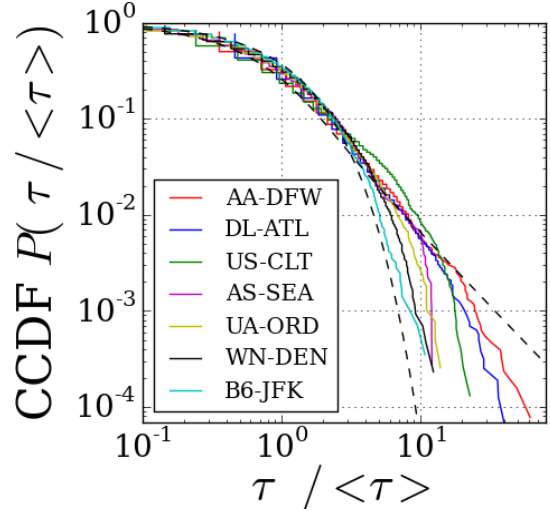}
\caption{(Color online) The CCDFs of the IETs of major airlines' airplanes in their main hub airports. Two dashed lines are the CCDFs of the IET of the Sine models for the parameters $a = 0.0$ (left) and $a = 1.0$ (right). The legends in the figure indicate airlines' IATA codes and their main hub airports. The IET distributions of these airlines' networks are heavy tailed compared with an exponential distribution while the extent of burstiness varies among these airlines. The IET distributions of most FSCs are characterized by power-law distributions with an exponent $\alpha=2.5$ and an exponential cutoff. In contrast, the IET distributions of LCCs are similar to an exponential distribution.}
\label{fig:6}
\end{center}
\end{figure}

The air transportation network is made from sub layers, and these sub layers are defined as each airline's network \cite{cardillo2013emergence}. The bursty behavior in an airport stems from each airline network's burstiness. We discuss the relationship between the airline networks and the extent of burstiness. We study the networks of seven major airlines in the U.S. and the extent of burstiness of the airplanes' arrival behavior in each airline's main hub airport (see Appendix A for a data description). First, we discuss the bursty behavior of arrivals operated by each airline in its hub airport. Figure~\ref{fig:6} shows the CCDFs of the IETs of the arrivals of the these airlines' airplanes at their main hub airports. The dashed lines on the left and right sides are the CCDFs of the IETs of Sine models for the parameters $a=0.0$ and $a=1.0$, respectively. The figure indicates that each IET distribution is heavy tailed in comparison with an exponential distribution and that some IET distributions follow a powerlaw distribution with an exponential cutoff. Each IET distribution varies in the cutoff values of its power-law tail. 

In Table~\ref{Table:2}, the basic properties of each airline network, the burst strength parameter, $a$, and the cutoff value, $\tau_0$, of the power-law distribution of the arrival behavior in its main hub airport are shown. The table shows that the five highest- and two lowest-ranking carriers in terms of the extent of burstiness are full-service carriers (FSCs) and low-cost carriers (LCCs), respectively. Only FSCs are characterized by the power-law tails of the IET distributions. This indicates that the types of carriers affect the extent of burstiness. 

\begin{figure}[t]
\begin{center}
\includegraphics[width=8.7cm]{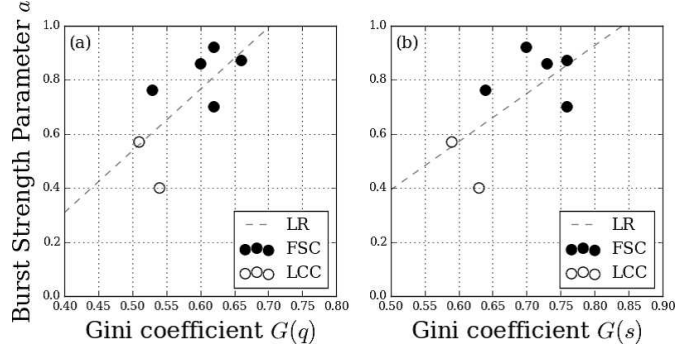}
\caption{The relationship between the degree Gini coefficient, $G(q)$, (a) and the strength Gini coefficient, $G(s)$, (b) and the burst strength parameter, $a$. LR represents the linear regression line. Both $G(q)$ and $G(s)$ are positively correlated with $a$.}
\label{fig:7}
\end{center}
\end{figure}

We investigate the reason for the difference in the extent of burstiness. The network structure of the air transportation system and the extent of the burstiness of its arrival behavior are strongly associated. The networks of LCCs are characterized by point-to-point networks, which are similar to the complete graph \cite{williams2001will}. In this type of network, the necessity of transit is low, since most nodes are connected with each other. By contrast, the networks of the FSCs follow a hub-and-spoke structure, which results from the necessity of the transit-facilitation strategy. Figure~\ref{fig:7} shows the relationships between the degree Gini coefficient, $G(q)$; the strength Gini coefficient, $G(s)$; and the burst strength parameter, $a$. The Gini coefficient can quantify the extent of the hub-and-spoke structure and accurately capture the characteristics of the FSCs' and LCCs' networks \cite{wuellner2010resilience}. The figure shows that the Gini coefficients and burst strength parameter are positively related, indicating that the larger the extent of the hub-and-spoke structure, the more bursty the arrival behavior. This figure supports the above reason as to why only the arrival behaviors of FSCs' airplanes are bursty. 

\section{DISCUSSION}
\label{sec:5}

In this paper, the burstiness of airplanes' arrival behavior in the air transportation system was analyzed. First, the empirical data of airplanes' arrival behavior in a wide range of hub airports were studied. It was universally observed that the CCDF of the IET in these airports followed power-law distributions with an exponent $\alpha = 2.5$ and an exponential cutoff. These also agreed well with the theoretically calculated IET distributions in the case of an inhomogeneous Poisson process whose event rate was given by $f(t)=Na \sin (2n \pi t) + 1~(0 \le t \le 1,~N \to \infty)$ (which is called the Sine model) regardless of differences in the locations and main carriers of airports. The extent of the burstiness quantified using the cutoff value of the power-law distribution and burst strength parameter was large in most airports. 

Moreover, the origin of the universally observed bursty behavior was investigated. Because of the network structure of the air transportation system (the so-called hub-and-spoke network with small-world and scale-free characteristics), the system has a strong demand to facilitate transit at hub airports. Passengers can easily transfer to connecting flights when airplanes arrive at the airports at almost the same time. This causes the number of airplanes' arrivals to fluctuate and the IET distribution to follow a power-law distribution with an exponent $\alpha = 2.5$ and an exponential cutoff. We verify this analysis by proposing Normal distribution model based on the mechanism mentioned above. This model is defined as an inhomogeneous Poisson process whose event rate is given by a mixture of normal distributions. Simulation and theoretical analysis of the model indicates that it can describe the bursty behavior observed in the empirical data. The mechanism is robust against the frequency of oscillation, the peakedness of the peaks, and the on-time performance of flights. This robustness contributes to the universality of bursty behavior in the air transportation system. 

Furthermore, we studied the relationship between each airline network and the bursty behavior of its arrivals at its main hub airport. The analysis indicated that the extent of the hub-and-spoke structure of airline networks and that of the burstiness of airplanes' arrivals were positively correlated. This result substantiated the mechanism for the bursty behavior described above. The hub-and-spoke airline networks of FSCs were characterized by transport of passengers from one peripheral airport to another via hub airports. Transit facilitation was necessary in these networks. In contrast, the extent of the hub-and-spoke structure of LCCs' networks was small, since most airports were connected by direct flights. Thus, transit facilitation was not necessary in these networks. The fact that the cutoff value was large in the case of airlines with hub-and-spoke networks indicated that transit facilitation played a key role in the burstiness of airplanes' arrival behavior. 

In conclusion, a universally observed bursty behavior was seen in the air transportation system, a human-made social network. Analyses on models and empirical data suggested that transit facilitation was the mechanism behind this behavior and that this mechanism was robust against variations of airports. The fact that many airports followed the same law was natural since, the system was optimized to maximize passengers' convenience and carriers' profit. In addition, the analysis of the necessity of transit facilitation by studying airline networks indicated that the bursty behavior originated from the hub-and-spoke network structure. One study has suggested that the heavy-tailed degree distribution of Wikipedia stems from bursty human activity \cite{muchnik2013origins}. These two results indicate that network characteristics, including small-world and scale-free, and activity behaviors such as burstiness are mutually correlated. The characteristic of one could not be fully understood without considering the other. Analysis on the relationship between these two is extremely important.

\section*{ACKNOWLEDGMENTS}
This work was supported by Grant-in-Aid for Scientific Research (No. 25287026) from the Japan Society for the Promotion of Science. We wish to thank Daichi Yanagisawa and Takahiro Ezaki for their valuable comments on this manuscript.

\section*{APPENDIX A: DATASET AND DATA PROCESSING}
\renewcommand{\theequation}{A.\arabic{equation}}
\setcounter{equation}{0}

The dataset in Sec~\ref{sec:2} shows the Airline On-Time Performance Data from the RITA database of the Bureau of Transportation Statistics (BTS) \cite{BTS}. Data from the 10 largest airports in the U.S. based on total passenger boarding in 2014 were used in Sec~\ref{sec:2}. The 10 airports were Hartsfield-Jackson Atlanta International Airport (ATL), Los Angeles International Airport (LAX), O'Hare International Airport (ORD), Dallas/Fort Worth International Airport (DFW), John F. Kennedy International Airport (JFK), Denver International Airport (DEN), San Francisco International Airport (SFO), Charlotte Douglas International Airport (CLT), McCarran International Airport (LAS), and Phoenix Sky Harbor International Airport (PHX). The IATA codes are written in parentheses. In addition, data from Seattle-Tacoma International Airport (SEA) were used in Sec~\ref{sec:4}. Moreover, the air transportation networks of the seven main airlines were analyzed in Sec~\ref{sec:4}. These airlines were American Airlines (AA), Delta Air Lines (DL), US Airways (US), Alaska Airlines (AS), United Airlines (UA), Southwest Airlines (WN), and JetBlue Airways (B6). The dataset contains on-time performance data such as scheduled and actual arrival times, destinations, and carriers for non-stop domestic flights. The data were reported by carriers with at least 1\% of the total domestic scheduled passenger revenue. The data from all flights operated in January 2014 were used in the analysis. The data from airplanes' arrivals and departures were recorded at 1 min intervals. The actual arrival times in the analysis were studied and canceled flights were removed from the IET distributions. The IETs across two business days were removed to exclude the influence of off-hours at night (see Appendix E for discussion on off-hours at night). The airport data in Table~\ref{Table:1} were also collected by BTS. The main carriers and their shares of airports were based on enplaned passengers (both arriving and departing).

\section*{APPENDIX B: VARIATION IN THE CUTOFF VALUE IN EACH AIRPORT}
\renewcommand{\theequation}{B.\arabic{equation}}
\setcounter{equation}{0}

We discuss two reasons why airports vary in the cutoff values listed in Table.~\ref{Table:1}. First, the fact that the LCCs' networks are characterized by small cutoff values mentioned in Sec~\ref{sec:4} explains the aggregated bursty behaviors in hub airports. In Table~\ref{Table:1}, WN and B6 are LCCs. The airport has a relatively small cutoff value if an LCC is a dominant carrier. Second, the main airline's share in each airport explains the extent of burstiness in airports. In Table~\ref{Table:1}, ATL, DFW, and CLT have quite large cutoff values. The main carriers in these airports have a high share. Each carrier concentrates airplanes' arrivals to facilitate transit; however, these carriers seldom cooperate unless they participate in the same alliance group. The amount of aggregated arrivals is a summation of each carrier's arrivals. If the majority of airplanes in an airport are operated by one carrier, the effect of concentration of arrivals because of the transit facilitation is strong, leading to a quite large cutoff value. By contrast, each carrier's share is not large in other airports. This weakens the effect of transit facilitation, leading to a relatively small cutoff value. 

In addition, JFK has a small burst strength parameter. This is because the available dataset is limited to only domestic flights. In JFK, the share of international flights in all scheduled flights is large. As a result, the bursty behavior in JFK is not appropriately assessed, which results in a small cutoff value.

\section*{APPENDIX C: DETAILS OF THE FITTING PROCESS OF THE MODEL TO THE EMPIRICAL DATA OF THE NORMAL DISTRIBUTION MODEL}
\renewcommand{\theequation}{C.\arabic{equation}}
\setcounter{equation}{0}
We discuss the fitting process in constructing the normal distribution model in Sec~\ref{sec:3}. We divide the time space of the empirical data into subregions and set a peak in each subregion as follows. The number of scheduled arrivals at a certain time of a day is counted. The data from different days are aggregated in counting for a sufficient amount of data. We define time as the minimum (maximum) time when the $t_{mv}$-min average of the number of scheduled arrivals at the time is minimal (maximal) in the range $t-t_{range} \le t \le \ t+t_{range}$. Let $t_{lmin}$s and $t_{lmax}$s denote the local minimum and maximum times, respectively. The first subregion $[t_{min,1},t_{max,1}]$ is defined as the region from $0$ to the minimum of $t_{lmin}$s and the $i$th subregion is defined by the region from $t_{max,i-1}$ to the minimum of $t_{lmin}$s such that $\exists t_{lmax}, t_{max,i-1} \le t_{lmax} \le t_{lmin}$. Then, the $i$th peak $t_{peak,i}$ is defined by the minimum of $t_{lmax}$s in the $i$th subregion. The parameters of the result in Fig~\ref{fig:4} are $t_{mv} = 5$ min and $t_{range} = 10$ min. In this case, the time space has 16 subregions.

\begin{figure}[t]
\begin{center}
\includegraphics[width=7cm]{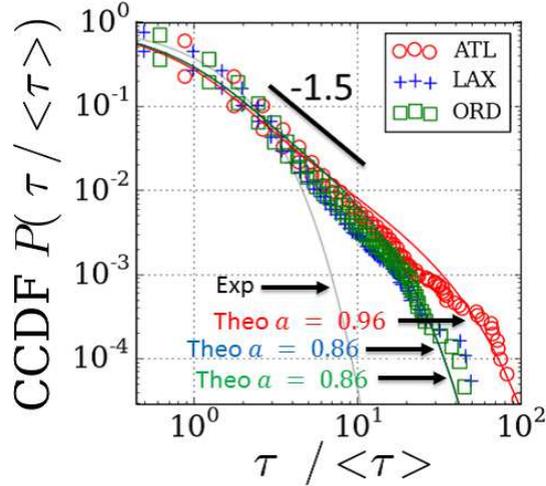}
\caption{(Color online) The CCDFs of the inter-departure time in three hub airports. The maximum and minimum of the probability are plotted for each IET $\tau$.}
\label{fig:8}
\end{center}
\end{figure}

\begin{figure}[t]
\begin{center}
\includegraphics[width=6cm]{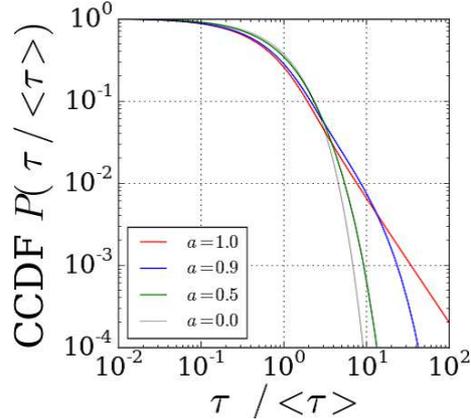}
\caption{(Color online) The theoretically calculated CCDFs of the IET of the Sine models for the parameters $a=0.0, 0.5, 0.9,$ and $1.0$. The event rate of the Sine model is given by $f(t)=N a \sin (2n \pi t) + 1~((0 \le t \le 1,~N \to \infty)$. When $a=1$, the IET distribution follows a power law with an exponent $\alpha=2.5$ without any cutoffs. Otherwise, the IET distribution follows a power law with an exponential cutoff. When $a=0$, the IET distribution is identical to the exponential distribution.}
\label{fig:2}
\end{center}
\end{figure}

\section*{APPENDIX D: BURSTINESS IN AIRPLANES' DEPARTURES}
\renewcommand{\theequation}{D.\arabic{equation}}
\setcounter{equation}{0}

The bursty behavior of airplanes' arrivals is studied in Sec.~\ref{sec:2}. In this Appendix, we study that of airplanes' departures. The CCDFs of the inter-departure time of airplanes in three hub airports are shown in Fig.~\ref{fig:8}. The distributions follow power laws with exponential cutoffs, just as inter-arrival time distributions do. This result indicates that airplanes' departure behavior also obeys the same mechanism generating burstiness.

However, compared with inter-arrival time distributions, the slopes of the inter-departure time distributions on a log-log plot are slightly steeper. This indicates that the event rate expands as $f(t - \tilde{t_i}) = c_{1i} + \mathcal{O}((t-\tilde{t_i})^n)$ asymptotically where $n<2$ at local minimum points if we assume that the airplanes' departures follow an inhomogeneous Poisson process \cite{vazquez2007impact1}. This result suggests that the delay distribution is not a smooth function because of artificial controls. It is difficult to artificially control the arrival times of airplanes, which results in the smooth event rate. By contrast, since airplanes are on the ground, it is relatively undemanding to control the waiting time before airplanes take off, especially when runways are not fully utilized. Since runways are likely to be vacant near the local minimum points, the effect of this artificial control makes the delay time distribution unsmooth when the event rate is low and the resulting $\tau$ is large.

\section*{APPENDIX E: EXISTENCE OF INACTIVE REGIONS IN EMPIRICAL DATA}
\renewcommand{\theequation}{E.\arabic{equation}}
\setcounter{equation}{0}

We discuss a model that follows an inhomogeneous Poisson process. If the event rate, $f(t)$, is positive, the IET converges to 0 in the limit $N \to \infty$, where $N$ is the average number of total events in a trial. However, if there is a region $t_{0,min} \le t \le t_{0,max}$ where $f(t) = 0$, the time interval between the last event before $t = t_{0,min}$ and the first event after $t = t_{0,max}$ is finite, even in the limit $N \to \infty$. Thus, the rescaled IET $\tau/\average{\tau}$ diverges to infinity in this limit. However, since the number of such regions is also finite, the percentage of the infinite rescaled IETs converges to 0 in this limit. Thus, it is not necessary to consider these regions if we take the limit. However, the number of events is finite in reality and these regions affect the result. We call the region with explicitly $f(t) \neq 0$ and $f(t) = 0$ as active and inactive regions, respectively. Most social behaviors have inactive regions, such as late at night when most people sleep. In the case of the air transportation system, no airplanes fly in most airports at night.

\section*{APPENDIX F: THEORETICAL ANALYSIS OF THE SINE MODEL}
\renewcommand{\theequation}{F.\arabic{equation}}
\setcounter{equation}{0}

We theoretically derive the CCDF of the IET of the Sine model in this Appendix. The event rate of the Sine model is $f(t)=N a \sin (2n \pi t) + 1~((0 \le t \le 1)$. In the limit $N \to \infty$, the event can be considered to occur every time. The rescaled IET at time $t$ is given by $\tau/\average{\tau} = X/f(t),$ where $X = N \Delta x$ is a stochastic variable whose PDF is $P(X) = e^{-X}$ and $1/f(t)$ represents the average event interval. Then, considering the whole time period, the distribution of $\tau/\average{\tau})$ is given by the product of the distributions of $X$ and $1/f(t)$. Then, we obtain
\begin{eqnarray}
\label{SinExact}
P(\tau/\average{\tau}) &=& \int_{0}^{2} \frac{a}{\pi} \left( \frac{\pi}{2} + \arcsin \left( x-1 \right) \right. \nonumber \\ &-& \left. a \sqrt{1-\left( x-1 \right)^2 } \right) \frac{\tau}{\average{\tau}} e^{-x \tau/\average{\tau}} dx \nonumber \\ &+& e^{-(1+a) \tau/\average{\tau}}.
\end{eqnarray} 
This result is independent of $n$ since the distribution of $1/f(t)$ is the same regardless of $n$. Moreover, when $\tau$ is sufficiently large, we obtain the approximate solution
\begin{eqnarray} 
\label{SinApprox}
P(\tau/\average{\tau}) &\sim& \frac{1+3a+8a(1-a)\tau/\average{\tau}}{8\sqrt{2\pi}a^{3/2}} \nonumber \\ &\times&  e^{-(1-a)\tau/\average{\tau}} \left(\frac{\tau}{\average{\tau}}\right)^{-3/2}.
\end{eqnarray} 
using Taylor expansion of functions from $x=0$, cutting off high-order terms, changing upper the limit of interval of integration in the first term to $\infty$, and ignoring the second term. Since the change in the value of $e^{-(1-a)\tau}[1+3a+8a(1-a)\tau]/8\sqrt{2\pi}a^{3/2}$ is small compared with $\tau^{-3/2}$ when $\tau < 1/(1-a)$, this result indicates that the CCDF of the IET of the Sine model follows a power-law function with exponent $\alpha= 5/2$ and a cutoff $\tau_0 = 1/(1-a)$.

The CCDF of the IET is shown in Fig.~\ref{fig:2}. A large parameter $a$ reflects a heavy-tailed IET distribution. The exponent of the power law is 2.5 and independent of the parameter $a$. However, the larger the parameter $a$, the larger the cutoff value of the IET distribution. When $a=0$, the IET distribution is given by an exponential distribution. In the case of $a>0$, the IET distribution is approximately given by a power-law distribution with an exponent $\alpha = 2.5$ and an exponential cutoff when $\tau$ is large. When $a=1$, the cutoff of the IET distribution vanishes. The parameters $a=0.5, 0.9$, and $1.0$ corresponds to the cutoff values $\tau_0 = 2.0, 10.0,$ and $\infty$. Any distributions between these extreme cases can be assessed using this parameter.


\begin{thebibliography}{49}%
\makeatletter
\providecommand \@ifxundefined [1]{%
 \@ifx{#1\undefined}
}%
\providecommand \@ifnum [1]{%
 \ifnum #1\expandafter \@firstoftwo
 \else \expandafter \@secondoftwo
 \fi
}%
\providecommand \@ifx [1]{%
 \ifx #1\expandafter \@firstoftwo
 \else \expandafter \@secondoftwo
 \fi
}%
\providecommand \natexlab [1]{#1}%
\providecommand \enquote  [1]{``#1''}%
\providecommand \bibnamefont  [1]{#1}%
\providecommand \bibfnamefont [1]{#1}%
\providecommand \citenamefont [1]{#1}%
\providecommand \href@noop [0]{\@secondoftwo}%
\providecommand \href [0]{\begingroup \@sanitize@url \@href}%
\providecommand \@href[1]{\@@startlink{#1}\@@href}%
\providecommand \@@href[1]{\endgroup#1\@@endlink}%
\providecommand \@sanitize@url [0]{\catcode `\\12\catcode `\$12\catcode
  `\&12\catcode `\#12\catcode `\^12\catcode `\_12\catcode `\%12\relax}%
\providecommand \@@startlink[1]{}%
\providecommand \@@endlink[0]{}%
\providecommand \url  [0]{\begingroup\@sanitize@url \@url }%
\providecommand \@url [1]{\endgroup\@href {#1}{\urlprefix }}%
\providecommand \urlprefix  [0]{URL }%
\providecommand \Eprint [0]{\href }%
\providecommand \doibase [0]{http://dx.doi.org/}%
\providecommand \selectlanguage [0]{\@gobble}%
\providecommand \bibinfo  [0]{\@secondoftwo}%
\providecommand \bibfield  [0]{\@secondoftwo}%
\providecommand \translation [1]{[#1]}%
\providecommand \BibitemOpen [0]{}%
\providecommand \bibitemStop [0]{}%
\providecommand \bibitemNoStop [0]{.\EOS\space}%
\providecommand \EOS [0]{\spacefactor3000\relax}%
\providecommand \BibitemShut  [1]{\csname bibitem#1\endcsname}%
\let\auto@bib@innerbib\@empty
%</preamble>
\bibitem [{\citenamefont {Albert}\ and\ \citenamefont
  {Barab\'asi}(2002)}]{RevModPhys.74.47}%
  \BibitemOpen
  \bibfield  {author} {\bibinfo {author} {\bibfnamefont {R.}~\bibnamefont
  {Albert}}\ and\ \bibinfo {author} {\bibfnamefont {A.-L.}\ \bibnamefont
  {Barab\'asi}},\ }\href {\doibase 10.1103/RevModPhys.74.47} {\bibfield
  {journal} {\bibinfo  {journal} {Rev. Mod. Phys.}\ }\textbf {\bibinfo {volume}
  {74}},\ \bibinfo {pages} {47} (\bibinfo {year} {2002})}\BibitemShut {NoStop}%
\bibitem [{\citenamefont {Castellano}\ \emph {et~al.}(2009)\citenamefont
  {Castellano}, \citenamefont {Fortunato},\ and\ \citenamefont
  {Loreto}}]{RevModPhys.81.591}%
  \BibitemOpen
  \bibfield  {author} {\bibinfo {author} {\bibfnamefont {C.}~\bibnamefont
  {Castellano}}, \bibinfo {author} {\bibfnamefont {S.}~\bibnamefont
  {Fortunato}}, \ and\ \bibinfo {author} {\bibfnamefont {V.}~\bibnamefont
  {Loreto}},\ }\href {\doibase 10.1103/RevModPhys.81.591} {\bibfield  {journal}
  {\bibinfo  {journal} {Rev. Mod. Phys.}\ }\textbf {\bibinfo {volume} {81}},\
  \bibinfo {pages} {591} (\bibinfo {year} {2009})}\BibitemShut {NoStop}%
\bibitem [{\citenamefont {Arenas}\ \emph {et~al.}(2008)\citenamefont {Arenas},
  \citenamefont {D{\'\i}az-Guilera}, \citenamefont {Kurths}, \citenamefont
  {Moreno},\ and\ \citenamefont {Zhou}}]{arenas2008synchronization}%
  \BibitemOpen
  \bibfield  {author} {\bibinfo {author} {\bibfnamefont {A.}~\bibnamefont
  {Arenas}}, \bibinfo {author} {\bibfnamefont {A.}~\bibnamefont
  {D{\'\i}az-Guilera}}, \bibinfo {author} {\bibfnamefont {J.}~\bibnamefont
  {Kurths}}, \bibinfo {author} {\bibfnamefont {Y.}~\bibnamefont {Moreno}}, \
  and\ \bibinfo {author} {\bibfnamefont {C.}~\bibnamefont {Zhou}},\ }\href@noop
  {} {\bibfield  {journal} {\bibinfo  {journal} {Phys. Rep.}\ }\textbf
  {\bibinfo {volume} {469}},\ \bibinfo {pages} {93} (\bibinfo {year}
  {2008})}\BibitemShut {NoStop}%
\bibitem [{\citenamefont {Barth{\'e}lemy}(2011)}]{barthelemy2011spatial}%
  \BibitemOpen
  \bibfield  {author} {\bibinfo {author} {\bibfnamefont {M.}~\bibnamefont
  {Barth{\'e}lemy}},\ }\href@noop {} {\bibfield  {journal} {\bibinfo  {journal}
  {Phys. Rep.}\ }\textbf {\bibinfo {volume} {499}},\ \bibinfo {pages} {1}
  (\bibinfo {year} {2011})}\BibitemShut {NoStop}%
\bibitem [{\citenamefont {Holme}\ and\ \citenamefont
  {Saram{\"a}ki}(2012)}]{holme2012temporal}%
  \BibitemOpen
  \bibfield  {author} {\bibinfo {author} {\bibfnamefont {P.}~\bibnamefont
  {Holme}}\ and\ \bibinfo {author} {\bibfnamefont {J.}~\bibnamefont
  {Saram{\"a}ki}},\ }\href@noop {} {\bibfield  {journal} {\bibinfo  {journal}
  {Phys. Rep.}\ }\textbf {\bibinfo {volume} {519}},\ \bibinfo {pages} {97}
  (\bibinfo {year} {2012})}\BibitemShut {NoStop}%
\bibitem [{\citenamefont {Eckmann}\ \emph {et~al.}(2004)\citenamefont
  {Eckmann}, \citenamefont {Moses},\ and\ \citenamefont
  {Sergi}}]{eckmann2004entropy}%
  \BibitemOpen
  \bibfield  {author} {\bibinfo {author} {\bibfnamefont {J.-P.}\ \bibnamefont
  {Eckmann}}, \bibinfo {author} {\bibfnamefont {E.}~\bibnamefont {Moses}}, \
  and\ \bibinfo {author} {\bibfnamefont {D.}~\bibnamefont {Sergi}},\
  }\href@noop {} {\bibfield  {journal} {\bibinfo  {journal} {Proc. Natl. Acad.
  Sci. USA}\ }\textbf {\bibinfo {volume} {101}},\ \bibinfo {pages} {14333}
  (\bibinfo {year} {2004})}\BibitemShut {NoStop}%
\bibitem [{\citenamefont {Karsai}\ \emph {et~al.}(2012)\citenamefont {Karsai},
  \citenamefont {Kaski}, \citenamefont {Barab{\'a}si},\ and\ \citenamefont
  {Kert{\'e}sz}}]{karsai2012universal}%
  \BibitemOpen
  \bibfield  {author} {\bibinfo {author} {\bibfnamefont {M.}~\bibnamefont
  {Karsai}}, \bibinfo {author} {\bibfnamefont {K.}~\bibnamefont {Kaski}},
  \bibinfo {author} {\bibfnamefont {A.-L.}\ \bibnamefont {Barab{\'a}si}}, \
  and\ \bibinfo {author} {\bibfnamefont {J.}~\bibnamefont {Kert{\'e}sz}},\
  }\href@noop {} {\bibfield  {journal} {\bibinfo  {journal} {Sci. Rep.}\
  }\textbf {\bibinfo {volume} {2}} (\bibinfo {year} {2012})}\BibitemShut
  {NoStop}%
\bibitem [{\citenamefont {V{\'a}zquez}\ \emph {et~al.}(2006)\citenamefont
  {V{\'a}zquez}, \citenamefont {Oliveira}, \citenamefont {Dezs{\"o}},
  \citenamefont {Goh}, \citenamefont {Kondor},\ and\ \citenamefont
  {Barab{\'a}si}}]{vazquez2006modeling}%
  \BibitemOpen
  \bibfield  {author} {\bibinfo {author} {\bibfnamefont {A.}~\bibnamefont
  {V{\'a}zquez}}, \bibinfo {author} {\bibfnamefont {J.~G.}\ \bibnamefont
  {Oliveira}}, \bibinfo {author} {\bibfnamefont {Z.}~\bibnamefont {Dezs{\"o}}},
  \bibinfo {author} {\bibfnamefont {K.-I.}\ \bibnamefont {Goh}}, \bibinfo
  {author} {\bibfnamefont {I.}~\bibnamefont {Kondor}}, \ and\ \bibinfo {author}
  {\bibfnamefont {A.-L.}\ \bibnamefont {Barab{\'a}si}},\ }\href@noop {}
  {\bibfield  {journal} {\bibinfo  {journal} {Phys. Rev. E}\ }\textbf {\bibinfo
  {volume} {73}},\ \bibinfo {pages} {036127} (\bibinfo {year}
  {2006})}\BibitemShut {NoStop}%
\bibitem [{\citenamefont {Goh}\ and\ \citenamefont
  {Barab{\'a}si}(2008)}]{goh2008burstiness}%
  \BibitemOpen
  \bibfield  {author} {\bibinfo {author} {\bibfnamefont {K.-I.}\ \bibnamefont
  {Goh}}\ and\ \bibinfo {author} {\bibfnamefont {A.-L.}\ \bibnamefont
  {Barab{\'a}si}},\ }\href@noop {} {\bibfield  {journal} {\bibinfo  {journal}
  {EPL}\ }\textbf {\bibinfo {volume} {81}},\ \bibinfo {pages} {48002} (\bibinfo
  {year} {2008})}\BibitemShut {NoStop}%
\bibitem [{\citenamefont {Candia}\ \emph {et~al.}(2008)\citenamefont {Candia},
  \citenamefont {Gonz{\'a}lez}, \citenamefont {Wang}, \citenamefont
  {Schoenharl}, \citenamefont {Madey},\ and\ \citenamefont
  {Barab{\'a}si}}]{candia2008uncovering}%
  \BibitemOpen
  \bibfield  {author} {\bibinfo {author} {\bibfnamefont {J.}~\bibnamefont
  {Candia}}, \bibinfo {author} {\bibfnamefont {M.~C.}\ \bibnamefont
  {Gonz{\'a}lez}}, \bibinfo {author} {\bibfnamefont {P.}~\bibnamefont {Wang}},
  \bibinfo {author} {\bibfnamefont {T.}~\bibnamefont {Schoenharl}}, \bibinfo
  {author} {\bibfnamefont {G.}~\bibnamefont {Madey}}, \ and\ \bibinfo {author}
  {\bibfnamefont {A.-L.}\ \bibnamefont {Barab{\'a}si}},\ }\href@noop {}
  {\bibfield  {journal} {\bibinfo  {journal} {J. Phys. A: Mathematical and
  Theoretical}\ }\textbf {\bibinfo {volume} {41}},\ \bibinfo {pages} {224015}
  (\bibinfo {year} {2008})}\BibitemShut {NoStop}%
\bibitem [{\citenamefont {Takaguchi}\ \emph {et~al.}(2011)\citenamefont
  {Takaguchi}, \citenamefont {Nakamura}, \citenamefont {Sato}, \citenamefont
  {Yano},\ and\ \citenamefont {Masuda}}]{takaguchi2011predictability}%
  \BibitemOpen
  \bibfield  {author} {\bibinfo {author} {\bibfnamefont {T.}~\bibnamefont
  {Takaguchi}}, \bibinfo {author} {\bibfnamefont {M.}~\bibnamefont {Nakamura}},
  \bibinfo {author} {\bibfnamefont {N.}~\bibnamefont {Sato}}, \bibinfo {author}
  {\bibfnamefont {K.}~\bibnamefont {Yano}}, \ and\ \bibinfo {author}
  {\bibfnamefont {N.}~\bibnamefont {Masuda}},\ }\href@noop {} {\bibfield
  {journal} {\bibinfo  {journal} {Phys. Rev. X}\ }\textbf {\bibinfo {volume}
  {1}},\ \bibinfo {pages} {011008} (\bibinfo {year} {2011})}\BibitemShut
  {NoStop}%
\bibitem [{\citenamefont {Cattuto}\ \emph {et~al.}(2010)\citenamefont
  {Cattuto}, \citenamefont {Van~den Broeck}, \citenamefont {Barrat},
  \citenamefont {Colizza}, \citenamefont {Pinton},\ and\ \citenamefont
  {Vespignani}}]{cattuto2010dynamics}%
  \BibitemOpen
  \bibfield  {author} {\bibinfo {author} {\bibfnamefont {C.}~\bibnamefont
  {Cattuto}}, \bibinfo {author} {\bibfnamefont {W.}~\bibnamefont {Van~den
  Broeck}}, \bibinfo {author} {\bibfnamefont {A.}~\bibnamefont {Barrat}},
  \bibinfo {author} {\bibfnamefont {V.}~\bibnamefont {Colizza}}, \bibinfo
  {author} {\bibfnamefont {J.-F.}\ \bibnamefont {Pinton}}, \ and\ \bibinfo
  {author} {\bibfnamefont {A.}~\bibnamefont {Vespignani}},\ }\href@noop {}
  {\bibfield  {journal} {\bibinfo  {journal} {PloS ONE}\ }\textbf {\bibinfo
  {volume} {5}},\ \bibinfo {pages} {e11596} (\bibinfo {year}
  {2010})}\BibitemShut {NoStop}%
\bibitem [{\citenamefont {Kemuriyama}\ \emph {et~al.}(2010)\citenamefont
  {Kemuriyama}, \citenamefont {Ohta}, \citenamefont {Sato}, \citenamefont
  {Maruyama}, \citenamefont {Tandai-Hiruma}, \citenamefont {Kato},\ and\
  \citenamefont {Nishida}}]{kemuriyama2010power}%
  \BibitemOpen
  \bibfield  {author} {\bibinfo {author} {\bibfnamefont {T.}~\bibnamefont
  {Kemuriyama}}, \bibinfo {author} {\bibfnamefont {H.}~\bibnamefont {Ohta}},
  \bibinfo {author} {\bibfnamefont {Y.}~\bibnamefont {Sato}}, \bibinfo {author}
  {\bibfnamefont {S.}~\bibnamefont {Maruyama}}, \bibinfo {author}
  {\bibfnamefont {M.}~\bibnamefont {Tandai-Hiruma}}, \bibinfo {author}
  {\bibfnamefont {K.}~\bibnamefont {Kato}}, \ and\ \bibinfo {author}
  {\bibfnamefont {Y.}~\bibnamefont {Nishida}},\ }\href@noop {} {\bibfield
  {journal} {\bibinfo  {journal} {BioSystems}\ }\textbf {\bibinfo {volume}
  {101}},\ \bibinfo {pages} {144} (\bibinfo {year} {2010})}\BibitemShut
  {NoStop}%
\bibitem [{\citenamefont {Saichev}\ and\ \citenamefont
  {Sornette}(2006)}]{saichev2006universal}%
  \BibitemOpen
  \bibfield  {author} {\bibinfo {author} {\bibfnamefont {A.}~\bibnamefont
  {Saichev}}\ and\ \bibinfo {author} {\bibfnamefont {D.}~\bibnamefont
  {Sornette}},\ }\href@noop {} {\bibfield  {journal} {\bibinfo  {journal}
  {Phys. Rev. Lett.}\ }\textbf {\bibinfo {volume} {97}},\ \bibinfo {pages}
  {078501} (\bibinfo {year} {2006})}\BibitemShut {NoStop}%
\bibitem [{\citenamefont {Moinet}\ \emph {et~al.}(2015)\citenamefont {Moinet},
  \citenamefont {Starnini},\ and\ \citenamefont
  {Pastor-Satorras}}]{PhysRevLett.114.108701}%
  \BibitemOpen
  \bibfield  {author} {\bibinfo {author} {\bibfnamefont {A.}~\bibnamefont
  {Moinet}}, \bibinfo {author} {\bibfnamefont {M.}~\bibnamefont {Starnini}}, \
  and\ \bibinfo {author} {\bibfnamefont {R.}~\bibnamefont {Pastor-Satorras}},\
  }\href {\doibase 10.1103/PhysRevLett.114.108701} {\bibfield  {journal}
  {\bibinfo  {journal} {Phys. Rev. Lett.}\ }\textbf {\bibinfo {volume} {114}},\
  \bibinfo {pages} {108701} (\bibinfo {year} {2015})}\BibitemShut {NoStop}%
\bibitem [{\citenamefont {Iribarren}\ and\ \citenamefont
  {Moro}(2009)}]{iribarren2009impact}%
  \BibitemOpen
  \bibfield  {author} {\bibinfo {author} {\bibfnamefont {J.~L.}\ \bibnamefont
  {Iribarren}}\ and\ \bibinfo {author} {\bibfnamefont {E.}~\bibnamefont
  {Moro}},\ }\href@noop {} {\bibfield  {journal} {\bibinfo  {journal} {Phys.
  Rev. Lett.}\ }\textbf {\bibinfo {volume} {103}},\ \bibinfo {pages} {038702}
  (\bibinfo {year} {2009})}\BibitemShut {NoStop}%
\bibitem [{\citenamefont {Gavald{\`a}-Miralles}\ \emph
  {et~al.}(2014)\citenamefont {Gavald{\`a}-Miralles}, \citenamefont {Choffnes},
  \citenamefont {Otto}, \citenamefont {S{\'a}nchez}, \citenamefont
  {Bustamante}, \citenamefont {Amaral}, \citenamefont {Duch},\ and\
  \citenamefont {Guimer{\`a}}}]{gavalda2014impact}%
  \BibitemOpen
  \bibfield  {author} {\bibinfo {author} {\bibfnamefont {A.}~\bibnamefont
  {Gavald{\`a}-Miralles}}, \bibinfo {author} {\bibfnamefont {D.~R.}\
  \bibnamefont {Choffnes}}, \bibinfo {author} {\bibfnamefont {J.~S.}\
  \bibnamefont {Otto}}, \bibinfo {author} {\bibfnamefont {M.~A.}\ \bibnamefont
  {S{\'a}nchez}}, \bibinfo {author} {\bibfnamefont {F.~E.}\ \bibnamefont
  {Bustamante}}, \bibinfo {author} {\bibfnamefont {L.~A.}\ \bibnamefont
  {Amaral}}, \bibinfo {author} {\bibfnamefont {J.}~\bibnamefont {Duch}}, \ and\
  \bibinfo {author} {\bibfnamefont {R.}~\bibnamefont {Guimer{\`a}}},\
  }\href@noop {} {\bibfield  {journal} {\bibinfo  {journal} {Proc. Natl. Acad.
  Sci. USA}\ }\textbf {\bibinfo {volume} {111}},\ \bibinfo {pages} {15322}
  (\bibinfo {year} {2014})}\BibitemShut {NoStop}%
\bibitem [{\citenamefont {Horv{\'a}th}\ and\ \citenamefont
  {Kert{\'e}sz}(2014)}]{horvath2014spreading}%
  \BibitemOpen
  \bibfield  {author} {\bibinfo {author} {\bibfnamefont {D.~X.}\ \bibnamefont
  {Horv{\'a}th}}\ and\ \bibinfo {author} {\bibfnamefont {J.}~\bibnamefont
  {Kert{\'e}sz}},\ }\href@noop {} {\bibfield  {journal} {\bibinfo  {journal}
  {New J. Phys.}\ }\textbf {\bibinfo {volume} {16}},\ \bibinfo {pages} {073037}
  (\bibinfo {year} {2014})}\BibitemShut {NoStop}%
\bibitem [{\citenamefont {Takaguchi}\ \emph {et~al.}(2013)\citenamefont
  {Takaguchi}, \citenamefont {Masuda},\ and\ \citenamefont
  {Holme}}]{10.1371/journal.pone.0068629}%
  \BibitemOpen
  \bibfield  {author} {\bibinfo {author} {\bibfnamefont {T.}~\bibnamefont
  {Takaguchi}}, \bibinfo {author} {\bibfnamefont {N.}~\bibnamefont {Masuda}}, \
  and\ \bibinfo {author} {\bibfnamefont {P.}~\bibnamefont {Holme}},\ }\href
  {\doibase 10.1371/journal.pone.0068629} {\bibfield  {journal} {\bibinfo
  {journal} {PLoS ONE}\ }\textbf {\bibinfo {volume} {8}},\ \bibinfo {pages}
  {e68629} (\bibinfo {year} {2013})}\BibitemShut {NoStop}%
\bibitem [{\citenamefont {Vazquez}\ \emph {et~al.}(2007)\citenamefont
  {Vazquez}, \citenamefont {Racz}, \citenamefont {Lukacs},\ and\ \citenamefont
  {Barab{\'a}si}}]{vazquez2007impact2}%
  \BibitemOpen
  \bibfield  {author} {\bibinfo {author} {\bibfnamefont {A.}~\bibnamefont
  {Vazquez}}, \bibinfo {author} {\bibfnamefont {B.}~\bibnamefont {Racz}},
  \bibinfo {author} {\bibfnamefont {A.}~\bibnamefont {Lukacs}}, \ and\ \bibinfo
  {author} {\bibfnamefont {A.-L.}\ \bibnamefont {Barab{\'a}si}},\ }\href@noop
  {} {\bibfield  {journal} {\bibinfo  {journal} {Phys. Rev. Lett.}\ }\textbf
  {\bibinfo {volume} {98}},\ \bibinfo {pages} {158702} (\bibinfo {year}
  {2007})}\BibitemShut {NoStop}%
\bibitem [{\citenamefont {Karsai}\ \emph {et~al.}(2011)\citenamefont {Karsai},
  \citenamefont {Kivel{\"a}}, \citenamefont {Pan}, \citenamefont {Kaski},
  \citenamefont {Kert{\'e}sz}, \citenamefont {Barab{\'a}si},\ and\
  \citenamefont {Saram{\"a}ki}}]{karsai2011small}%
  \BibitemOpen
  \bibfield  {author} {\bibinfo {author} {\bibfnamefont {M.}~\bibnamefont
  {Karsai}}, \bibinfo {author} {\bibfnamefont {M.}~\bibnamefont {Kivel{\"a}}},
  \bibinfo {author} {\bibfnamefont {R.~K.}\ \bibnamefont {Pan}}, \bibinfo
  {author} {\bibfnamefont {K.}~\bibnamefont {Kaski}}, \bibinfo {author}
  {\bibfnamefont {J.}~\bibnamefont {Kert{\'e}sz}}, \bibinfo {author}
  {\bibfnamefont {A.-L.}\ \bibnamefont {Barab{\'a}si}}, \ and\ \bibinfo
  {author} {\bibfnamefont {J.}~\bibnamefont {Saram{\"a}ki}},\ }\href@noop {}
  {\bibfield  {journal} {\bibinfo  {journal} {Phys. Rev. E}\ }\textbf {\bibinfo
  {volume} {83}},\ \bibinfo {pages} {025102} (\bibinfo {year}
  {2011})}\BibitemShut {NoStop}%
\bibitem [{\citenamefont {Jo}\ \emph {et~al.}(2014)\citenamefont {Jo},
  \citenamefont {Perotti}, \citenamefont {Kaski},\ and\ \citenamefont
  {Kert{\'e}sz}}]{jo2014analytically}%
  \BibitemOpen
  \bibfield  {author} {\bibinfo {author} {\bibfnamefont {H.-H.}\ \bibnamefont
  {Jo}}, \bibinfo {author} {\bibfnamefont {J.~I.}\ \bibnamefont {Perotti}},
  \bibinfo {author} {\bibfnamefont {K.}~\bibnamefont {Kaski}}, \ and\ \bibinfo
  {author} {\bibfnamefont {J.}~\bibnamefont {Kert{\'e}sz}},\ }\href@noop {}
  {\bibfield  {journal} {\bibinfo  {journal} {Phys. Rev. X}\ }\textbf {\bibinfo
  {volume} {4}},\ \bibinfo {pages} {011041} (\bibinfo {year}
  {2014})}\BibitemShut {NoStop}%
\bibitem [{\citenamefont {Hidalgo}\ and\ \citenamefont
  {C{\'e}sar}(2006)}]{hidalgo2006conditions}%
  \BibitemOpen
  \bibfield  {author} {\bibinfo {author} {\bibfnamefont {R.}~\bibnamefont
  {Hidalgo}}\ and\ \bibinfo {author} {\bibfnamefont {A.}~\bibnamefont
  {C{\'e}sar}},\ }\href@noop {} {\bibfield  {journal} {\bibinfo  {journal}
  {Physica A}\ }\textbf {\bibinfo {volume} {369}},\ \bibinfo {pages} {877}
  (\bibinfo {year} {2006})}\BibitemShut {NoStop}%
\bibitem [{\citenamefont {Barab{\'a}si}(2005)}]{barabasi2005origin}%
  \BibitemOpen
  \bibfield  {author} {\bibinfo {author} {\bibfnamefont {A.-L.}\ \bibnamefont
  {Barab{\'a}si}},\ }\href@noop {} {\bibfield  {journal} {\bibinfo  {journal}
  {Nature (London)}\ }\textbf {\bibinfo {volume} {435}},\ \bibinfo {pages}
  {207} (\bibinfo {year} {2005})}\BibitemShut {NoStop}%
\bibitem [{\citenamefont {Vazquez}(2005)}]{vazquez2005exact}%
  \BibitemOpen
  \bibfield  {author} {\bibinfo {author} {\bibfnamefont {A.}~\bibnamefont
  {Vazquez}},\ }\href@noop {} {\bibfield  {journal} {\bibinfo  {journal} {Phys.
  Rev. Lett.}\ }\textbf {\bibinfo {volume} {95}},\ \bibinfo {pages} {248701}
  (\bibinfo {year} {2005})}\BibitemShut {NoStop}%
\bibitem [{\citenamefont {Vajna}\ \emph {et~al.}(2013)\citenamefont {Vajna},
  \citenamefont {T{\'o}th},\ and\ \citenamefont
  {Kert{\'e}sz}}]{vajna2013modelling}%
  \BibitemOpen
  \bibfield  {author} {\bibinfo {author} {\bibfnamefont {S.}~\bibnamefont
  {Vajna}}, \bibinfo {author} {\bibfnamefont {B.}~\bibnamefont {T{\'o}th}}, \
  and\ \bibinfo {author} {\bibfnamefont {J.}~\bibnamefont {Kert{\'e}sz}},\
  }\href@noop {} {\bibfield  {journal} {\bibinfo  {journal} {New J. Phys.}\
  }\textbf {\bibinfo {volume} {15}},\ \bibinfo {pages} {103023} (\bibinfo
  {year} {2013})}\BibitemShut {NoStop}%
\bibitem [{\citenamefont {Malmgren}\ \emph {et~al.}(2008)\citenamefont
  {Malmgren}, \citenamefont {Stouffer}, \citenamefont {Motter},\ and\
  \citenamefont {Amaral}}]{malmgren2008poissonian}%
  \BibitemOpen
  \bibfield  {author} {\bibinfo {author} {\bibfnamefont {R.~D.}\ \bibnamefont
  {Malmgren}}, \bibinfo {author} {\bibfnamefont {D.~B.}\ \bibnamefont
  {Stouffer}}, \bibinfo {author} {\bibfnamefont {A.~E.}\ \bibnamefont
  {Motter}}, \ and\ \bibinfo {author} {\bibfnamefont {L.~A.}\ \bibnamefont
  {Amaral}},\ }\href@noop {} {\bibfield  {journal} {\bibinfo  {journal} {Proc.
  Natl. Acad. Sci. USA}\ }\textbf {\bibinfo {volume} {105}},\ \bibinfo {pages}
  {18153} (\bibinfo {year} {2008})}\BibitemShut {NoStop}%
\bibitem [{\citenamefont {Jo}\ \emph {et~al.}(2012)\citenamefont {Jo},
  \citenamefont {Karsai}, \citenamefont {Kert{\'e}sz},\ and\ \citenamefont
  {Kaski}}]{jo2012circadian}%
  \BibitemOpen
  \bibfield  {author} {\bibinfo {author} {\bibfnamefont {H.-H.}\ \bibnamefont
  {Jo}}, \bibinfo {author} {\bibfnamefont {M.}~\bibnamefont {Karsai}}, \bibinfo
  {author} {\bibfnamefont {J.}~\bibnamefont {Kert{\'e}sz}}, \ and\ \bibinfo
  {author} {\bibfnamefont {K.}~\bibnamefont {Kaski}},\ }\href@noop {}
  {\bibfield  {journal} {\bibinfo  {journal} {New J. Phys.}\ }\textbf {\bibinfo
  {volume} {14}},\ \bibinfo {pages} {013055} (\bibinfo {year}
  {2012})}\BibitemShut {NoStop}%
\bibitem [{\citenamefont {Masuda}\ \emph {et~al.}(2013)\citenamefont {Masuda},
  \citenamefont {Takaguchi}, \citenamefont {Sato},\ and\ \citenamefont
  {Yano}}]{masuda2013self}%
  \BibitemOpen
  \bibfield  {author} {\bibinfo {author} {\bibfnamefont {N.}~\bibnamefont
  {Masuda}}, \bibinfo {author} {\bibfnamefont {T.}~\bibnamefont {Takaguchi}},
  \bibinfo {author} {\bibfnamefont {N.}~\bibnamefont {Sato}}, \ and\ \bibinfo
  {author} {\bibfnamefont {K.}~\bibnamefont {Yano}},\ }in\ \href@noop {} {\emph
  {\bibinfo {booktitle} {Temporal Networks}}}\ (\bibinfo  {publisher}
  {Springer},\ \bibinfo {year} {2013})\ pp.\ \bibinfo {pages}
  {245--264}\BibitemShut {NoStop}%
\bibitem [{\citenamefont {Vestergaard}\ \emph {et~al.}(2014)\citenamefont
  {Vestergaard}, \citenamefont {G{\'e}nois},\ and\ \citenamefont
  {Barrat}}]{vestergaard2014memory}%
  \BibitemOpen
  \bibfield  {author} {\bibinfo {author} {\bibfnamefont {C.~L.}\ \bibnamefont
  {Vestergaard}}, \bibinfo {author} {\bibfnamefont {M.}~\bibnamefont
  {G{\'e}nois}}, \ and\ \bibinfo {author} {\bibfnamefont {A.}~\bibnamefont
  {Barrat}},\ }\href@noop {} {\bibfield  {journal} {\bibinfo  {journal} {Phys.
  Rev. E}\ }\textbf {\bibinfo {volume} {90}},\ \bibinfo {pages} {042805}
  (\bibinfo {year} {2014})}\BibitemShut {NoStop}%
\bibitem [{\citenamefont {Vazquez}(2007)}]{vazquez2007impact1}%
  \BibitemOpen
  \bibfield  {author} {\bibinfo {author} {\bibfnamefont {A.}~\bibnamefont
  {Vazquez}},\ }\href@noop {} {\bibfield  {journal} {\bibinfo  {journal}
  {Physica A}\ }\textbf {\bibinfo {volume} {373}},\ \bibinfo {pages} {747}
  (\bibinfo {year} {2007})}\BibitemShut {NoStop}%
\bibitem [{\citenamefont {Bryan}\ and\ \citenamefont
  {O'Kelly}(1999)}]{bryan1999hub}%
  \BibitemOpen
  \bibfield  {author} {\bibinfo {author} {\bibfnamefont {D.~L.}\ \bibnamefont
  {Bryan}}\ and\ \bibinfo {author} {\bibfnamefont {M.~E.}\ \bibnamefont
  {O'Kelly}},\ }\href@noop {} {\bibfield  {journal} {\bibinfo  {journal} {J.
  Reg. Sci.}\ }\textbf {\bibinfo {volume} {39}},\ \bibinfo {pages} {275}
  (\bibinfo {year} {1999})}\BibitemShut {NoStop}%
\bibitem [{\citenamefont {Guimera}\ \emph {et~al.}(2005)\citenamefont
  {Guimera}, \citenamefont {Mossa}, \citenamefont {Turtschi},\ and\
  \citenamefont {Amaral}}]{guimera2005worldwide}%
  \BibitemOpen
  \bibfield  {author} {\bibinfo {author} {\bibfnamefont {R.}~\bibnamefont
  {Guimera}}, \bibinfo {author} {\bibfnamefont {S.}~\bibnamefont {Mossa}},
  \bibinfo {author} {\bibfnamefont {A.}~\bibnamefont {Turtschi}}, \ and\
  \bibinfo {author} {\bibfnamefont {L.~N.}\ \bibnamefont {Amaral}},\
  }\href@noop {} {\bibfield  {journal} {\bibinfo  {journal} {Proc. Natl. Acad.
  Sci. USA}\ }\textbf {\bibinfo {volume} {102}},\ \bibinfo {pages} {7794}
  (\bibinfo {year} {2005})}\BibitemShut {NoStop}%
\bibitem [{\citenamefont {Zanin}\ and\ \citenamefont
  {Lillo}(2013)}]{zanin2013modelling}%
  \BibitemOpen
  \bibfield  {author} {\bibinfo {author} {\bibfnamefont {M.}~\bibnamefont
  {Zanin}}\ and\ \bibinfo {author} {\bibfnamefont {F.}~\bibnamefont {Lillo}},\
  }\href@noop {} {\bibfield  {journal} {\bibinfo  {journal} {Eur. Phys. J.
  Special Topics}\ }\textbf {\bibinfo {volume} {215}},\ \bibinfo {pages} {5}
  (\bibinfo {year} {2013})}\BibitemShut {NoStop}%
\bibitem [{\citenamefont {Li}\ and\ \citenamefont
  {Cai}(2004)}]{li2004statistical}%
  \BibitemOpen
  \bibfield  {author} {\bibinfo {author} {\bibfnamefont {W.}~\bibnamefont
  {Li}}\ and\ \bibinfo {author} {\bibfnamefont {X.}~\bibnamefont {Cai}},\
  }\href@noop {} {\bibfield  {journal} {\bibinfo  {journal} {Phys. Rev. E}\
  }\textbf {\bibinfo {volume} {69}},\ \bibinfo {pages} {046106} (\bibinfo
  {year} {2004})}\BibitemShut {NoStop}%
\bibitem [{\citenamefont {Cardillo}\ \emph {et~al.}(2013)\citenamefont
  {Cardillo}, \citenamefont {G{\'o}mez-Garde{\~n}es}, \citenamefont {Zanin},
  \citenamefont {Romance}, \citenamefont {Papo}, \citenamefont {del Pozo},\
  and\ \citenamefont {Boccaletti}}]{cardillo2013emergence}%
  \BibitemOpen
  \bibfield  {author} {\bibinfo {author} {\bibfnamefont {A.}~\bibnamefont
  {Cardillo}}, \bibinfo {author} {\bibfnamefont {J.}~\bibnamefont
  {G{\'o}mez-Garde{\~n}es}}, \bibinfo {author} {\bibfnamefont {M.}~\bibnamefont
  {Zanin}}, \bibinfo {author} {\bibfnamefont {M.}~\bibnamefont {Romance}},
  \bibinfo {author} {\bibfnamefont {D.}~\bibnamefont {Papo}}, \bibinfo {author}
  {\bibfnamefont {F.}~\bibnamefont {del Pozo}}, \ and\ \bibinfo {author}
  {\bibfnamefont {S.}~\bibnamefont {Boccaletti}},\ }\href@noop {} {\bibfield
  {journal} {\bibinfo  {journal} {Sci. Rep.}\ }\textbf {\bibinfo {volume} {3}}
  (\bibinfo {year} {2013})}\BibitemShut {NoStop}%
\bibitem [{\citenamefont {Colizza}\ \emph {et~al.}(2006)\citenamefont
  {Colizza}, \citenamefont {Barrat}, \citenamefont {Barth{\'e}lemy},\ and\
  \citenamefont {Vespignani}}]{colizza2006role}%
  \BibitemOpen
  \bibfield  {author} {\bibinfo {author} {\bibfnamefont {V.}~\bibnamefont
  {Colizza}}, \bibinfo {author} {\bibfnamefont {A.}~\bibnamefont {Barrat}},
  \bibinfo {author} {\bibfnamefont {M.}~\bibnamefont {Barth{\'e}lemy}}, \ and\
  \bibinfo {author} {\bibfnamefont {A.}~\bibnamefont {Vespignani}},\
  }\href@noop {} {\bibfield  {journal} {\bibinfo  {journal} {Proc. Natl. Acad.
  Sci. USA}\ }\textbf {\bibinfo {volume} {103}},\ \bibinfo {pages} {2015}
  (\bibinfo {year} {2006})}\BibitemShut {NoStop}%
\bibitem [{\citenamefont {Wuellner}\ \emph {et~al.}(2010)\citenamefont
  {Wuellner}, \citenamefont {Roy},\ and\ \citenamefont
  {D'fSouza}}]{wuellner2010resilience}%
  \BibitemOpen
  \bibfield  {author} {\bibinfo {author} {\bibfnamefont {D.~R.}\ \bibnamefont
  {Wuellner}}, \bibinfo {author} {\bibfnamefont {S.}~\bibnamefont {Roy}}, \
  and\ \bibinfo {author} {\bibfnamefont {R.~M.}\ \bibnamefont {D'fSouza}},\
  }\href@noop {} {\bibfield  {journal} {\bibinfo  {journal} {Phys. Rev. E}\
  }\textbf {\bibinfo {volume} {82}},\ \bibinfo {pages} {056101} (\bibinfo
  {year} {2010})}\BibitemShut {NoStop}%
\bibitem [{\citenamefont {Fleurquin}\ \emph {et~al.}(2013)\citenamefont
  {Fleurquin}, \citenamefont {Ramasco},\ and\ \citenamefont
  {Eguiluz}}]{fleurquin2013systemic}%
  \BibitemOpen
  \bibfield  {author} {\bibinfo {author} {\bibfnamefont {P.}~\bibnamefont
  {Fleurquin}}, \bibinfo {author} {\bibfnamefont {J.~J.}\ \bibnamefont
  {Ramasco}}, \ and\ \bibinfo {author} {\bibfnamefont {V.~M.}\ \bibnamefont
  {Eguiluz}},\ }\href@noop {} {\bibfield  {journal} {\bibinfo  {journal} {Sci.
  Rep.}\ }\textbf {\bibinfo {volume} {3}} (\bibinfo {year} {2013})}\BibitemShut
  {NoStop}%
\bibitem [{\citenamefont {Peterson}\ \emph {et~al.}(1995)\citenamefont
  {Peterson}, \citenamefont {Bertsimas},\ and\ \citenamefont
  {Odoni}}]{peterson1995models}%
  \BibitemOpen
  \bibfield  {author} {\bibinfo {author} {\bibfnamefont {M.~D.}\ \bibnamefont
  {Peterson}}, \bibinfo {author} {\bibfnamefont {D.~J.}\ \bibnamefont
  {Bertsimas}}, \ and\ \bibinfo {author} {\bibfnamefont {A.~R.}\ \bibnamefont
  {Odoni}},\ }\href@noop {} {\bibfield  {journal} {\bibinfo  {journal}
  {Manag. Sci.}\ }\textbf {\bibinfo {volume} {41}},\ \bibinfo {pages}
  {1279} (\bibinfo {year} {1995})}\BibitemShut {NoStop}%
\bibitem [{\citenamefont {Gwiggner}\ and\ \citenamefont
  {Nagaoka}(2014)}]{gwiggner2014data}%
  \BibitemOpen
  \bibfield  {author} {\bibinfo {author} {\bibfnamefont {C.}~\bibnamefont
  {Gwiggner}}\ and\ \bibinfo {author} {\bibfnamefont {S.}~\bibnamefont
  {Nagaoka}},\ }\href@noop {} {\bibfield  {journal} {\bibinfo  {journal} {Eur.
  J. Oper. Res.}\ }\textbf {\bibinfo {volume} {235}},\ \bibinfo {pages} {265}
  (\bibinfo {year} {2014})}\BibitemShut {NoStop}%
\bibitem [{\citenamefont {Galliher}\ and\ \citenamefont
  {Wheeler}(1958)}]{galliher1958nonstationary}%
  \BibitemOpen
  \bibfield  {author} {\bibinfo {author} {\bibfnamefont {H.~P.}\ \bibnamefont
  {Galliher}}\ and\ \bibinfo {author} {\bibfnamefont {R.~C.}\ \bibnamefont
  {Wheeler}},\ }\href@noop {} {\bibfield  {journal} {\bibinfo  {journal}
  {Oper. Res.}\ }\textbf {\bibinfo {volume} {6}},\ \bibinfo {pages}
  {264} (\bibinfo {year} {1958})}\BibitemShut {NoStop}%
\bibitem [{\citenamefont {Pan}\ and\ \citenamefont
  {Saram{\"a}ki}(2011)}]{pan2011path}%
  \BibitemOpen
  \bibfield  {author} {\bibinfo {author} {\bibfnamefont {R.~K.}\ \bibnamefont
  {Pan}}\ and\ \bibinfo {author} {\bibfnamefont {J.}~\bibnamefont
  {Saram{\"a}ki}},\ }\href@noop {} {\bibfield  {journal} {\bibinfo  {journal}
  {Phys. Rev. E}\ }\textbf {\bibinfo {volume} {84}},\ \bibinfo {pages} {016105}
  (\bibinfo {year} {2011})}\BibitemShut {NoStop}%
\bibitem [{\citenamefont {Barnhart}\ \emph {et~al.}(2014)\citenamefont
  {Barnhart}, \citenamefont {Fearing},\ and\ \citenamefont
  {Vaze}}]{barnhart2014modeling}%
  \BibitemOpen
  \bibfield  {author} {\bibinfo {author} {\bibfnamefont {C.}~\bibnamefont
  {Barnhart}}, \bibinfo {author} {\bibfnamefont {D.}~\bibnamefont {Fearing}}, \
  and\ \bibinfo {author} {\bibfnamefont {V.}~\bibnamefont {Vaze}},\ }\href@noop
  {} {\bibfield  {journal} {\bibinfo  {journal} {Oper. Res.}\ }\textbf
  {\bibinfo {volume} {62}},\ \bibinfo {pages} {580} (\bibinfo {year}
  {2014})}\BibitemShut {NoStop}%
\bibitem [{\citenamefont {Mueller}\ and\ \citenamefont
  {Chatterji}(2002)}]{mueller2002analysis}%
  \BibitemOpen
  \bibfield  {author} {\bibinfo {author} {\bibfnamefont {E.~R.}\ \bibnamefont
  {Mueller}}\ and\ \bibinfo {author} {\bibfnamefont {G.~B.}\ \bibnamefont
  {Chatterji}},\ }in\ \href@noop {} {\emph {\bibinfo {booktitle} {AIAA aircraft
  technology, integration and operations (ATIO) conference}}}\ (\bibinfo {year}
  {2002})\BibitemShut {NoStop}%
\bibitem [{\citenamefont {Lan}\ \emph {et~al.}(2006)\citenamefont {Lan},
  \citenamefont {Clarke},\ and\ \citenamefont {Barnhart}}]{lan2006planning}%
  \BibitemOpen
  \bibfield  {author} {\bibinfo {author} {\bibfnamefont {S.}~\bibnamefont
  {Lan}}, \bibinfo {author} {\bibfnamefont {J.-P.}\ \bibnamefont {Clarke}}, \
  and\ \bibinfo {author} {\bibfnamefont {C.}~\bibnamefont {Barnhart}},\
  }\href@noop {} {\bibfield  {journal} {\bibinfo  {journal} {Transportation
  Sci.}\ }\textbf {\bibinfo {volume} {40}},\ \bibinfo {pages} {15} (\bibinfo
  {year} {2006})}\BibitemShut {NoStop}%
\bibitem [{\citenamefont {Williams}(2001)}]{williams2001will}%
  \BibitemOpen
  \bibfield  {author} {\bibinfo {author} {\bibfnamefont {G.}~\bibnamefont
  {Williams}},\ }\href@noop {} {\bibfield  {journal} {\bibinfo  {journal} {J.
  Air Transp. Manag.}\ }\textbf {\bibinfo {volume} {7}},\ \bibinfo {pages}
  {277} (\bibinfo {year} {2001})}\BibitemShut {NoStop}%
\bibitem [{\citenamefont {Muchnik}\ \emph {et~al.}(2013)\citenamefont
  {Muchnik}, \citenamefont {Pei}, \citenamefont {Parra}, \citenamefont {Reis},
  \citenamefont {Andrade~Jr}, \citenamefont {Havlin},\ and\ \citenamefont
  {Makse}}]{muchnik2013origins}%
  \BibitemOpen
  \bibfield  {author} {\bibinfo {author} {\bibfnamefont {L.}~\bibnamefont
  {Muchnik}}, \bibinfo {author} {\bibfnamefont {S.}~\bibnamefont {Pei}},
  \bibinfo {author} {\bibfnamefont {L.~C.}\ \bibnamefont {Parra}}, \bibinfo
  {author} {\bibfnamefont {S.~D.}\ \bibnamefont {Reis}}, \bibinfo {author}
  {\bibfnamefont {J.~S.}\ \bibnamefont {Andrade~Jr}}, \bibinfo {author}
  {\bibfnamefont {S.}~\bibnamefont {Havlin}}, \ and\ \bibinfo {author}
  {\bibfnamefont {H.~A.}\ \bibnamefont {Makse}},\ }\href@noop {} {\bibfield
  {journal} {\bibinfo  {journal} {Sci. Rep.}\ }\textbf {\bibinfo {volume} {3}}
  (\bibinfo {year} {2013})}\BibitemShut {NoStop}%
\bibitem [{\citenamefont {of~Transportation Statistics~(BTS)}(2014)}]{BTS}%
  \BibitemOpen
  \bibfield  {author} {\bibinfo {author} {\bibfnamefont {Bureau}~\bibnamefont
  {of~Transportation Statistics~(BTS)}},\ }\href {http://www.rita.dot.gov/bts/}
  {\enquote {\bibinfo {title} {Rita database},}\ } (\bibinfo {year}
  {2014}).

\end{thebibliography}
\end{document}